\documentclass[12pt,draftclsnofoot,journal,onecolumn]{IEEEtran}

\usepackage[dvips]{graphicx}
\usepackage{float,makeidx,cite}
\usepackage{amssymb,amsmath,amsthm}

\providecommand{\abs}[1]{\lvert#1\rvert}
\newcommand{\ud}{\,\mathrm{d}}

\newtheorem{Propi1}{Proposition}
\newtheorem{Propi2}[Propi1]{Proposition}

\newtheorem{Theoi1}{Theorem}
\newtheorem{Theoi2}[Theoi1]{Theorem}

\title{Downlink Coordinated Joint Transmission for Mutual Information Accumulation}
\author{Amogh Rajanna, \IEEEmembership{Member, IEEE}, and Martin Haenggi, \IEEEmembership{Fellow, IEEE}
\thanks{A. Rajanna and M. Haenggi are with the
Wireless Institute, University of Notre Dame, USA. Email \{arajanna,mhaenggi\}@nd.edu.}}

\begin{document}
\maketitle
\begin{abstract}
In this letter, we propose a new coordinated multipoint (CoMP) technique based on mutual information (MI) accumulation using rateless codes. Using a stochastic geometry model for the cellular downlink, we quantify the performance enhancements in coverage probability and rate due to MI accumulation. By simulation and analysis, we show that MI accumulation using rateless codes leads to remarkable improvements in coverage and rate for general users and specific cell edge users.
\end{abstract}

\begin{IEEEkeywords}
CoMP, Cloud RAN, Rateless Codes, 5G Cellular Downlink, Stochastic Geometry, PPP, Joint Transmission.
\end{IEEEkeywords}
\section{Introduction}
Cloud radio access based cellular networks are envisioned to be based on \emph{amorphous} architectures rather than a strict cell-based design.
In an amorphous cloud RAN downlink setting, a user will be served by more than one nearest BS through joint transmission. In this letter, we propose a new CoMP technique where the coordinating BSs jointly transmit multiple codewords of the same information packet using rateless codes \cite{JourVer,AmoHae} leading to MI accumulation at the user. Modeling the BS locations by a Poisson point process (PPP), we provide expressions for the success (coverage) probability and rate of the typical user under the proposed CoMP scheme. We show that with MI accumulation using rateless codes, the users observe high coverage benefit and the following rate gains: the users close to \emph{only one} BS and the users equidistant from \emph{three} BSs have a rate increase by a factor of $2.6$ and $6.12$, respectively.
\section{MI Accumulation} 
\label{sec:CodeComb}
The cooperating BSs jointly transmit a $K$-bit information packet to the user. These BSs have access to the $K$ bits through the X2/S1 interface of backhaul connection to the cloud. Each transmitting BS encodes the $K$-bit packet with a unique rateless code, i.e., if two cooperating BSs transmit to a user, then they encode the $K$-bit packet with two different rateless codes and transmit the two codewords incrementally. In order for the user to resolve (distinguish) the multiple codewords successfully, the codewords can be communicated over orthogonal frequency bands. The cooperating BSs can also use NOMA schemes such as sparse code multiple access, lattice partition multiple access etc to transmit the codewords to the user with added receiver complexity \cite{NOMA, Molisch}. The sum rate in the MAC capacity region is achieved by the cooperating BSs. The multiple codewords are combined at the user in an iterative decoder to decode the $K$ bits. This decoding process leads to MI accumulation at the user. The achievable rate  at the user is given by 
\begin{equation}
C=\sum_{i=1}^M \log_2\left(1+\mathrm{SIR}_i\right), \label{C_exp}
\end{equation}
where $\log_2\left(1+\mathrm{SIR}_i\right)$ is the MI of codeword $i$ from cooperating BS $i$. From now onwards, $M=1$ and $M>1$ are referred to as the no cooperation (NC) and MI accumulation (MIA) schemes, respectively. In this letter, we derive results for $M=1$ and $M=2$.

\section{Performance Characterization}
\label{sec:Appli}
\subsection{System Model}
\label{sysMod}
We consider two independent homogeneous PPPs $\Phi_1$ and $\Phi_2$ of intensity $\lambda/2$. The nodes in $\Phi_k$ represent BSs using exclusively spreading code $k\in\{1,2\}$. We assume a single tier cellular downlink in which BSs are modeled by a PPP $\Phi=\Phi_1\cup\Phi_2=\{X_i\},~i=1,2,\cdots$. 
The typical user is located at the origin. The distance between the typical user and BS $X_i$  is $D_i$. Each BS uses constant transmit power. The channel is quasi-static flat fading affected by path loss. The typical user receives a $K$-bit packet from the nearest one or more cooperating BSs. During the typical user reception time, we assume that the interfering BSs are transmitting \emph{continuously}. 

The interference power and SIR at the typical user based on only the nearest BS $X_1$ transmission are given by
\begin{equation}
I_1=\sum_{k\neq 1}\abs{h_{k}}^2|X_k|^{-\alpha}
\label{int_eq}
\end{equation}
\begin{equation}
\mathrm{SIR}_1=\frac{\abs{h_{1}}^2D_1^{-\alpha}
}{I_1}\label{sir_in}.
\end{equation}
Each packet transmission of $K$ bits has a delay constraint of $N$ channel uses. The time to decode a $K$-bit packet $\hat{T}$ and the packet transmission time $T$ are given by
\begin{align}
&\hat{T}=\min\left\{t:K<t\cdot C\right\}\label{Rx_pkt}\\
&T=\min\left(N,\hat{T}\right).\label{pkt_Ti}
\end{align}
The packet transmission time $T$ quantifies the benefits of MIA using rateless codes as a CoMP scheme. 

The two metrics used to compare the performance of CoMP schemes in this letter are the success probability and rate of $K$-bit packet transmission, defined as 
\begin{align}
p_s(N)&\triangleq 1-\mathbb{P}\left(\hat T> N\right)\label{p_s}\\
R_N&\triangleq \frac{Kp_s(N)}{\mathbb{E}\left[T\right]}.\label{Rn}		
\end{align}
Both $p_s(N)$ and $R_N$ depend on the distribution of $T$. 
Based on (\ref{pkt_Ti}), the CCDF of $T$ is 
$\mathbb{P}\left(T>t\right)=\mathbb{P}(\hat{T}>t)$, $t<N$. The CCDF of $\hat T$ is given by
\begin{align}
A_M\left(t\right)&\triangleq \mathbb{P}(\hat{T}>t)
=\mathbb{P}\Big (K/t\geq \sum_{i=1}^M\log_2\left(1+\mathrm{SIR}_i\right)\Big).\label{ccdf_eq}
\end{align}
In the following, we consider two classes of users, the typical general user and the typical worst-case user, in the single tier cellular downlink and discuss their packet transmission time distributions.
\subsection{General User}
\label{casGU}
The typical general user is located at the origin, without conditioning on $\Phi$. Its performance corresponds to the spatial average over all users in $\mathbb{R}^2$. For the NC case, the user is served only by the nearest BS and the performance is determined by the $\mathrm{SIR}$ distribution\cite{AndrewsTC,Nigam}
\begin{equation}
\mathbb{P}\left(\mathrm{SIR}>\nu\right)\triangleq G(\nu)=\frac{1}{{}_2F_{1}\left(\left[1,-\delta\right]; 1-\delta;-\nu\right)}\label{Gdf}.
\end{equation}
where ${}_2F_{1}\left([a, b]; c; z\right)$ is the Gauss hypergeometric function and $\delta=2/\alpha$. 
\begin{Propi2}
\label{Pro2}
For the general user without cooperation, the CCDF of $T$ is given by
\begin{align}
\mathbb{P}\left(T>t\right)&=1-G(2^{K/t}-1),\quad t<N\label{GuNC}.
\end{align}
\end{Propi2}
\begin{IEEEproof}
The CCDF is obtained by computing $A_1\left(t\right)$ in (\ref{ccdf_eq}) for the general user.
\end{IEEEproof}
In the case of MIA, the user is served by the nearest BS in both $\Phi_1$ and $\Phi_2$. The two cooperating BSs transmit $2$ codewords on $2$ different spreading codes. Also the two BSs have \emph{i.i.d distances} to the typical user. 
\begin{Theoi1}
\label{The1}
The CCDF of the general user packet transmission time with MI accumulation, $T$ in (\ref{pkt_Ti}), is lower bounded as
\begin{align}
&\mathbb{P}\left(T>t\right)\geq \int_0^{\gamma} \left(G\left(\gamma-y\right)-1\right)G'(y)\ud y, \label{MIpktG}
\end{align}
where $\gamma=2\left(2^{K/2t}-1\right)$ and $G(y)$ is defined in (\ref{Gdf}) with its derivative $G'(y)$ based on the derivative of the hypergeometric function
\begin{equation}
	\frac{\ud}{\ud y} {}_2F_{1}\left(\left[a,b\right];c;y\right)=\frac{a b}{c} {}_2F_{1}\left(\left[a+1,b+1\right];c+1;y\right).
\end{equation}
\end{Theoi1}
\begin{IEEEproof}
The proof is obtained by evaluating (\ref{ccdf_eq}) with $M=2$ for the general user. Using the arithmetic-geometric mean inequality, a bound on (\ref{ccdf_eq}) can be obtained. Let $Y_i=\mathrm{SIR}_i$,
\begin{align}
	\sqrt{\left(1+Y_1\right)\left(1+Y_2\right)}&\leq 1+\frac{1}{2}\left(Y_1+Y_2\right)\Rightarrow\nonumber\\
	\sum \log_2\left(1+Y_i\right)&\leq 2 \log_2\left(1+\frac{1}{2}\left(Y_1+Y_2\right)\right). \label{SE_ub}
\end{align}

Using (\ref{SE_ub}), a lower bound for (\ref{ccdf_eq}) is given by
\begin{equation}
A_2\left(t\right) \geq \mathbb{P}\left(Y_1+Y_2 \leq 2\left(2^{K/2t}-1\right)\right).\label{Gu_LB}
\end{equation}

Note that $Y_1$ and $Y_2$ are i.i.d. To evaluate (\ref{Gu_LB}), we need the CCDF of $Y_i$, i.e., $\bar F_{Y_i}(y)=G(y)$ given in (\ref{Gdf}). Then the bound in (\ref{Gu_LB}) can be computed based on
\begin{align}
\mathbb{P}\left(Y_1+Y_2 \leq \gamma\right)&=\mathbb{E}\left[\mathbb{P}\left(Y_1 \leq \gamma-Y_2\mid Y_2\right)\right]\nonumber\\
&=-\int_0^{\gamma} \left(1-\bar F_{Y_1}(\gamma-y)\right) \ud \bar F_{Y_2}(y).\label{In_eq}
\end{align}
\end{IEEEproof}
From (\ref{p_s}) and (\ref{ccdf_eq}), the success probability is given by
\begin{equation}
p_s(N)=1-A_M\left(N\right).\label{ps_Fex}
\end{equation}
For the NC and MIA schemes, the $A_1\left(N\right)$ and $A_2\left(N\right)$ are given in (\ref{GuNC}) and (\ref{MIpktG}) at $t=N$ respectively. The rate is given by
\begin{equation}
R_N= \frac{Kp_s(N)}{\int_0^N A_M\left(t\right) \ud t}.\label{Rn_Fex}
\end{equation}
Now we quantify the performance benefits of MIA as a CoMP scheme relative to the NC.

The definition of diversity gain of a CoMP scheme with fixed-rate coding appears in \cite{Nigam}. Equivalently, for rateless coding, the diversity gain is defined as
\begin{align}
g_{\rm d}&\triangleq \lim_{N\rightarrow \infty} \frac{\log\left( 1-p_s(N)\right)}{-\log N}\label{DGexp}.
\end{align}
Based on (\ref{ps_Fex}), the diversity gain can be obtained by quantifying the scaling of outage probability $A_M(N)$ as $N\rightarrow \infty$. $\theta_N= 2^{K/N}-1$ has the following Taylor series%
\begin{equation}
\theta_N=\frac{K\log 2}{N}+ O\left(\frac{1}{N^2}\right),\quad N\rightarrow \infty\label{TS_the}.
\end{equation}
For the NC case, the scaling of $A_1(N)$ is obtained by combining (\ref{TS_the}) with the asymptotic result from \cite[Sec II.B]{Haen}
\begin{align}
&\mathbb{P}\left(Y_1 \leq y\right)\sim y\frac{\delta}{1-\delta},\quad y\rightarrow 0 \label{Y1asy}\\
&A_1(N)=\mathbb{P}\left(Y_1 \leq \theta_N\right) \sim \frac{K\log 2}{N}\frac{\delta}{1-\delta},~N\rightarrow \infty \label{NCscal}.
\end{align}
For the MIA scheme, the scaling of $A_2(N)$ is computed below. For i.i.d $Y_1$ and $Y_2$, using (\ref{In_eq}) and (\ref{Y1asy}), 
it can be shown that
\begin{equation}
\mathbb{P}\left(Y_1+Y_2 \leq y\right)\sim \left(\frac{\delta}{1-\delta}\right)^2 \frac{y^2}{2},\quad y \rightarrow 0.\label{A2Nexp}
\end{equation}
Hence
\begin{align}
&A_2\left(N\right) \geq \mathbb{P}\left(Y_1+Y_2 \leq 2\left(2^{K/2N}-1\right)\right)\nonumber\\
&\quad \stackrel{(a)}{\sim} \frac{1}{2}\left(\frac{\delta}{1-\delta}\right)^2 \left(\frac{K\log 2}{N}\right)^2,~N\rightarrow \infty, \label{MIscal}
\end{align}
where (a) follows by applying the Taylor series to $2(2^{K/2N}-1)$ similar to (\ref{TS_the}). Based on (\ref{DGexp}) and (\ref{NCscal}), it is observed that $g_{\rm d}=1$ for NC. For MIA, 
\begin{align}
g_{\rm d}&=\lim_{N\rightarrow \infty} \frac{2\log\left(\frac{\delta}{1-\delta}\frac{K\log 2}{N\sqrt{2}}\right)}{-\log N}=2\label{DG_mia}.
\end{align}
The diversity gain $g_{\rm d}$ provides the scaling law of the outage probability, i.e., rate of decay as $N\rightarrow \infty$. Thus, the scaling of the outage probability of MIA can be interpreted as $A_2(N)\sim A_1(N)^2/2$. 
\subsection{Worst-case user}
\label{casWU}
Letting $\mathcal{V}$ denote the set of Voronoi vertices of the BS PPP $\Phi=\Phi_1\cup\Phi_2$ (which is itself a stationary point process),
the typical worst-case user is obtained by placing a user at the origin $o$ and conditioning $\Phi$ on $o\in\mathcal{V}$. The worst-case user performance corresponds to the average over all points in $\mathcal{V}$, which are equidistant from three BSs.  

In the Voronoi tessellation of $\Phi$, a worst-case user will have $3$ equidistant BSs sharing either the same spreading code or two unique spreading codes with probability $0.25$ and $0.75$, respectively. In the first case, there will be no MI accumulation but amplitude accumulation of the transmitted signals\cite{Nigam}. In the latter case, the user achieves MI accumulation. 

For the NC scheme, the user is served by only one of the three equidistant BSs. The two other equidistant BSs together with the further away BSs in the network act as interferers. 
\begin{Propi1}
\label{Pro1}
For the worst-case user without cooperation ($M=1$), the CCDF of $T$ is given by
\begin{align}
\mathbb{P}\left(T>t\right)&=1-\left[\frac{1/\left(1+\theta\right)}{{}_2F_{1}\left(\left[1,-\delta\right]; 1-\delta;-\theta\right)}\right]^2, \label{WuNC}
\end{align}
where $\theta=2^{K/t}-1$.
\end{Propi1}
\begin{IEEEproof}
The proof is based on computing $A_1\left(t\right)$ in (\ref{ccdf_eq}) for the worst-case user. For  details, refer to \cite[Appendix~A]{AmoHae} (which pertains to the general user case). Let $D$ be the distance of the typical user from the $3$ equidistant BSs in $\Phi$. Its pdf is given by $f_D\left(r\right)=\exp\left(-\pi\lambda r^2\right)\left(\pi\lambda\right)^2 r^3$ from \cite{DiPap}.
\begin{align}
A_1\left(t\right)&=\mathbb{P}\left(\theta\geq \frac{\abs{h_{1}}^2D^{-\alpha}}{\left(\abs{h_2}^2+\abs{h_3}^2\right)D^{-\alpha}+I_1}\right)\label{A1Wu}\\
&\stackrel{(a)}{=}1-\mathbb{E}\left[\exp\left(-\abs{h_2}^2-\abs{h_3}^2-D^{\alpha}I_1\right)\theta\mid D\right]\nonumber\\
&\stackrel{(b)}{=}1-\frac{1}{\left(1+\theta\right)^2} \mathbb{E}\left[\exp\left(-\pi \lambda H(\theta) D^2\right)\right]\nonumber\\
&\stackrel{(c)}{=}1-\frac{1}{\left(1+\theta\right)^2} \frac{1}{\left(1+H(\theta)\right)^2},\label{Pro2Eqn}
\end{align} 
where (a) is due to $\abs{h_1}^2\sim$ Exp($1$), (b) follows from the Laplace transform (LT) of two Exp($1$) RVs at $\theta$, the LT of $I_1$ at $\theta D^{\alpha}$ and $H(\theta)=\frac{\delta \theta}{1-\delta}~{}_2F_{1} \left(\left[1,1-\delta\right];2-\delta;-\theta\right)$. The $\mathbb{E}\left[\cdot\right]$ w.r.t the pdf of $D$ leads to (c). The hypergeometric identity $1+H(\theta)={}_2F_{1}\left(\left[1,-\delta\right]; 1-\delta;-\theta\right)$ yields (\ref{WuNC}) from (\ref{Pro2Eqn}).
\end{IEEEproof}
For the MIA scheme, we assume that each user is served by the two equidistant BSs with unique spreading codes ($M=2$) with the third equidistant BS interfering.
\begin{Theoi2}
\label{The2}
The CCDF of the worst-case user packet transmission time with MI accumulation, $T$ in (\ref{pkt_Ti}), is lower bounded as
\begin{align}
&\mathbb{P}\left(T>t\right)\geq \int_0^{\infty}\!\!\!\int_0^{\gamma} \left(U\left(\gamma-y\right)-1\right)\tilde{G}'(y)f_D(r)\ud y \ud r\label{MIpktW}\\
&\tilde{G}(y)=\exp\left(-\pi\frac{\lambda}{2}r^2\left({}_2F_{1}\left(
\left[1,-\delta\right];1-\delta;-y\right)-1\right)\right)\label{G_qtn}\\
&U(y)=\frac{\tilde G(y)}{1+y},\quad \gamma=2\left(2^{K/2t}-1\right)\label{DdfWu}.
\end{align}
\end{Theoi2}
\begin{IEEEproof}
To compute the CCDF of $T$ for the worst-case user as per (\ref{ccdf_eq}) with $M=2$, we note that $\mathrm{SIR}_i$, $i\in \{1,2\}$ are both dependent on $D$. For $\mathrm{SIR}_1$, there will be one interfering BS at distance $D$. The other interfering BSs are further away. For $\mathrm{SIR}_2$, all interferers are further away than distance $D$. Hence to evaluate (\ref{ccdf_eq}), we define two RVs $\tilde Y_1$ and $\tilde Y_2$ similar to Section \ref{casGU}
\begin{align}
\tilde Y_1&=\frac{\abs{h_1}^2r^{-\alpha}}{\abs{h_3}^2r^{-\alpha}+I_1},\quad \tilde Y_2=\frac{\abs{h_{2}}^2r^{-\alpha}}{I_2} \label{Y1_eq},
\end{align}
where $r$ is the sample value of $D$. Note that $I_1$ and $I_2$ are from two independent PPPs of intensity $\lambda/2$. Then (\ref{ccdf_eq}) can be written as
\begin{align}
A_2\left(t\right)&=\int \mathbb{P}\big (K/t>\sum_{i=1}^2\log_2\left(1+\tilde Y_i\right)\big)f_D\left(r\right)\ud r.\label{Wu_ev}
\end{align}
Using (\ref{SE_ub}), a lower bound for (\ref{Wu_ev}) is given by
\begin{equation}
A_2\left(t\right) \geq \int \mathbb{P}\left(\tilde Y_1+\tilde Y_2 \leq 2\left(2^{K/2t}-1\right)\right)f_D\left(r\right)\ud r.\label{Wu_LB}
\end{equation}

The CCDFs of $\tilde Y_1$ and $\tilde Y_2$ are given by
\begin{align}
\bar F_{\tilde Y_2}(y)&=\mathbb{P}\left(\tilde Y_2\geq y\right)=\tilde G(y) \label{Y2df}\\
\bar F_{\tilde Y_1}(y)&=U\left(y\right)=\frac{\tilde G\left(y\right)}{1+y},\label{Y1df}
\end{align}
where $\tilde G(\cdot)$ is defined in (\ref{G_qtn}). The proof follows steps quite similar to Proposition \ref{Pro1}, (\ref{A1Wu})-(\ref{Pro2Eqn}) except for deterministic $r$. 
Note that $\tilde Y_1$ and $\tilde Y_2$ are independent and hence, (\ref{In_eq}) can be used to further evaluate (\ref{Wu_LB}), completing the proof. 
\end{IEEEproof}
Based on (\ref{WuNC}) and (\ref{MIpktW}), the resulting $p_s(N)$ and $R_N$ can be computed numerically for the worst-case user. In the following, we derive the diversity gain result for the worst-case user.

$A_1(N)$ is obtained from (\ref{A1Wu}) with $\theta$ replaced by $\theta_N$. Using similar arguments as in (\ref{TS_the})-(\ref{NCscal}), we get
\begin{equation}
A_1(N)\sim \frac{K\log 2}{N}\left(2+\frac{2\delta}{1-\delta}\right),\quad N\rightarrow \infty \label{NCscaWu}.
\end{equation}
Comparing (\ref{NCscal}) and (\ref{NCscaWu}), the additive $2$ appears due to the two equidistant interferers and the function of $\delta$ is scaled by $2$ since the worst-case user is further away from the serving BS than the general user, i.e., the mean distance of the worst-case user to the serving BS is $50\%$ more than that of general user. Hence from (\ref{DGexp}), $g_{\rm d}=1$ for NC. 

The scaling of $A_2(N)$ is obtained based on (\ref{Wu_LB}). For $\tilde Y_1$ and $\tilde Y_2$ in (\ref{Y1_eq}), the asymptotic CDF is obtained based on the scaling behavior of $\tilde G(y)$ in (\ref{G_qtn}) as $y\rightarrow 0$
\begin{align}
\mathbb{P}\left(\tilde Y_2< y\right)&=1-\tilde G(y)\stackrel{(a)}{\sim} \pi\frac{\lambda}{2}r^2\frac{\delta}{1-\delta}y, \quad y\rightarrow 0\label{Y2_sca}\\
\mathbb{P}\left(\tilde Y_1< y\right)&=1-U(y)\sim 1-\tilde G(y),\quad y\rightarrow 0\label{Y1_sca}.
\end{align}
where (a) follows from $e^{-x}\sim 1-x$, $x\rightarrow 0$ and using the previously used hypergeometric identity along with ${}_2F_{1}\left(\left[1,1-\delta\right];2-\delta;-x\right)\rightarrow 1$ as $x\rightarrow 0$. Using (\ref{Y2_sca}) and (\ref{Y1_sca}), the asymptotic CDF of $\tilde Y_1+\tilde Y_2$ can be obtained similar to (\ref{A2Nexp}). Hence the scaling of $A_2(N)$ is given by
\begin{equation}
A_2(N)\sim \left(\frac{K\log 2}{N\sqrt{2}}\right)^2 \int \Big(\pi\frac{\lambda}{2}r^2\frac{\delta}{1-\delta}\Big)^2 f_D\left(r\right) \ud r, ~ N\rightarrow \infty. \nonumber	
\end{equation}
From above, it follows that $g_{\rm d}=2$ for MIA as per (\ref{DGexp}).
\section{Numerical Results}
\label{sec:Num_Results}
\begin{figure}[!hbtp]
\centering
\includegraphics[scale=0.55, width=0.5\textwidth]{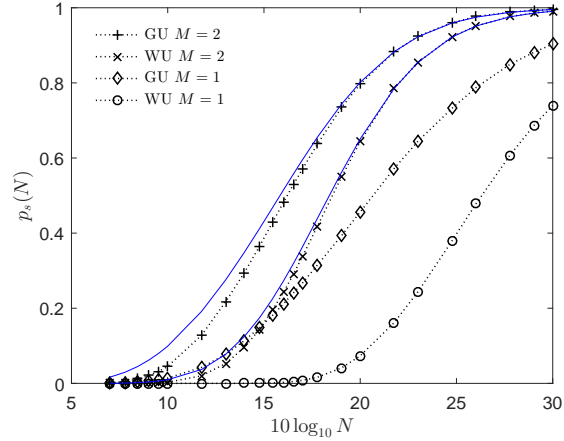}
\caption{The success probability $p_s(N)$ as a function of the delay constraint $N$ from (\ref{p_s}) and (\ref{ps_Fex}). GU-General user and WU-Worst user. For MIA, the dotted line is simulation based and the solid lines are obtained from the analytical results.}
\label{Psucc_vsN}
\end{figure}
In Fig. \ref{Psucc_vsN}, a plot of the success probability against the delay constraint $N$ is shown for a cellular network with $\lambda=1$ at $\alpha=3$ and $K=75$ bits. In MIA, both worst-case and general users have reduced interference on each codeword compared to the NC case due to the presence of unique spreading codes. In the NC scheme, the worst-case user has $2$ interfering BSs at the same distance as the desired BS and the general user has all interferers further away than desired BS. In the MIA scheme, the worst-case user has two cooperating BSs at the same distance whereas the two cooperating BSs of general user are at i.i.d. distances. Hence the worst-case user benefit represents the best possible coverage improvement due to MIA.
\begin{figure}[!hbtp]
\centering
\includegraphics[scale=0.55, width=0.5\textwidth]{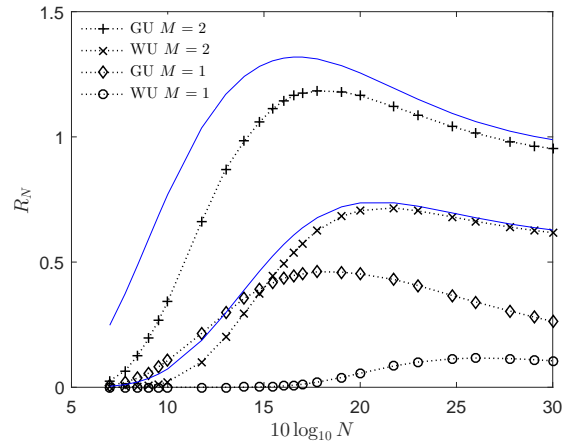}
\caption{The typical user rate $R_N$ as a function of $N$ from (\ref{Rn}) and (\ref{Rn_Fex}).} 
\label{Rate_vsN}
\end{figure}

Fig. \ref{Rate_vsN} shows a plot of the rate $R_N$ against the delay constraint $N$ for a cellular network with $\lambda=1$ at $\alpha=3$. While the $p_s(N)$ curves in Fig. \ref{Psucc_vsN} show the benefit due to MIA \emph{only}, the $R_N$ curves in Fig. \ref{Rate_vsN} show the benefits of both MIA and rateless coding. The effect of rateless coding is captured in the $R_N$ expression by the term $\mathbb{E}\left[T\right]$ (for fixed rate coding, this term is fixed to $N$). The rate gain $g_{\rm r}$ is defined as \emph{the ratio of the maximal rate $\max_N R_N$ with MIA to that with NC}. The $R_N$ curves for the worst-case user show a rate gain $g_{\rm r}$ of $6.12$, while for the general user curves, the rate gain $g_{\rm r}$ is $2.6$. Since rateless codes adapt to the instantaneous channel conditions, the replacement of an interfering BS by a cooperating BS leads to a big decrease in $\mathbb{E}\left[T\right]$ (more for worst-case user) yielding a higher rate as per (\ref{Rn}). The rate benefit of MIA is further enhanced by using the adaptive rateless codes. Although not considered in this work, for users on the Voronoi edges of the PPP $\Phi$, i.e., users equidistant from $2$ BSs, the rate gain $g_{\rm r}$ is expected to be between $2.6$ and $6.12$. 

\textbf{Cost of NOMA}: One cost of NOMA is the inter-codeword interference cancellation module at the receiver. Since $M=2$ codewords are multiplexed by the NOMA scheme in this paper, the hardware complexity needed to distinguish the two codewords at the receiver is affordable. A second cost is the excess bandwidth factor $\beta$ needed for NOMA. $\beta=2$ is a reasonable assumption for $M=2$. (Note that orthogonal frequency bands also require $\beta=2$). To have a net gain in the cost-benefit tradeoff of the new CoMP scheme, the rate gain $g_{\rm r}$ should satisfy $g_{\rm r}>\beta$. Both the general and worst-case (also cell edge) users satisfy $g_{\rm r}>\beta$ and thus have a net gain.
\section{Conclusion}
\label{sec:Concl}
We introduced a new CoMP scheme leveraging the effects of spectral efficiency boosting MI accumulation and the channel adaptivity of rateless codes. The resulting performance improvements are illustrated for a single tier cellular downlink for representative network scenarios. The users closer to the interfering BSs experience the most coverage and rate benefits. The presented CoMP scheme can be incorporated into a 2-tier (or $M$-tier) cellular model. For a user, the nearest BS in each tier performs joint transmission of the two codewords. 
\bibliography{References_PD}

\begin{thebibliography}{1}
\providecommand{\url}[1]{#1}
\csname url@samestyle\endcsname
\providecommand{\newblock}{\relax}
\providecommand{\bibinfo}[2]{#2}
\providecommand{\BIBentrySTDinterwordspacing}{\spaceskip=0pt\relax}
\providecommand{\BIBentryALTinterwordstretchfactor}{4}
\providecommand{\BIBentryALTinterwordspacing}{\spaceskip=\fontdimen2\font plus
\BIBentryALTinterwordstretchfactor\fontdimen3\font minus
  \fontdimen4\font\relax}
\providecommand{\BIBforeignlanguage}[2]{{%
\expandafter\ifx\csname l@#1\endcsname\relax
\typeout{** WARNING: IEEEtran.bst: No hyphenation pattern has been}%
\typeout{** loaded for the language `#1'. Using the pattern for}%
\typeout{** the default language instead.}%
\else
\language=\csname l@#1\endcsname
\fi
#2}}
\providecommand{\BIBdecl}{\relax}
\BIBdecl

\bibitem{JourVer}
A.~Rajanna, I.~Bergel, and M.~Kaveh, ``{Performance Analysis of Rateless Codes
  in an ALOHA Wireless Adhoc Network},'' \emph{IEEE Transactions on Wireless
  Communications}, vol.~14, no.~11, pp. 6216--6229, Nov 2015.

\bibitem{AmoHae}
A.~Rajanna and M.~Haenggi, ``{Enhanced Cellular Coverage and Throughput using
  Rateless Codes},'' \emph{IEEE Transactions on Communications}, Jul 2016,
  submitted for publication (http://arxiv.org/abs/1608.00269).

\bibitem{NOMA}
Y.~Saito, Y.~Kishiyama, A.~Benjebbour, T.~Nakamura, A.~Li, and K.~Higuchi,
  ``{Non-Orthogonal Multiple Access (NOMA) for Cellular Future Radio Access},''
  in \emph{2013 IEEE 77th Vehicular Technology Conference (VTC Spring)}, June
  2013, pp. 1--5.

\bibitem{Molisch}
A.~F. Molisch, N.~B. Mehta, J.~Yedidia, and J.~Zhang, ``{Performance of
  Fountain codes in Collaborative Relay Networks},'' \emph{IEEE Transactions on
  Wireless Communications}, vol.~6, no.~11, pp. 4108--4119, Nov 2007.

\bibitem{AndrewsTC}
J.~G. Andrews, F.~Baccelli, and R.~K. Ganti, ``{A tractable approach to
  coverage and rate in cellular networks},'' \emph{IEEE Transactions on
  Communications}, vol.~59, no.~11, pp. 3122--3134, Nov 2011.

\bibitem{Nigam}
G.~Nigam, P.~Minero, and M.~Haenggi, ``{Coordinated Multipoint Joint
  Transmission in Heterogeneous Networks},'' \emph{IEEE Transactions on
  Communications}, vol.~62, no.~11, pp. 4134--4146, Nov 2014.

\bibitem{Haen}
M.~Haenggi, ``{The Mean Interference-to-Signal Ratio and Its Key Role in
  Cellular and Amorphous Networks},'' \emph{IEEE Wireless Communications
  Letters}, vol.~3, no.~6, pp. 597--600, Dec 2014.

\bibitem{DiPap}
V.~Baumstark and G.~Last, ``{Some distributional results for Poisson- Voronoi
  tessellations},'' \emph{Advances Appl. Probability}, vol.~39, no.~1, pp.
  16--40, Mar 2007.

\end{thebibliography}
\bibliographystyle{IEEEtran}
\end{document}